\documentclass[showpacs, prepintnumber,amssymb, nobibnotes, aps, prd]{revtex4}
\usepackage{graphicx}
\usepackage{dcolumn}
\usepackage{bm}

\begin{document}

\title{S-WAVE SCATTERING OF FERMION REVISITED} 

\author{Anisur Rahaman}
\email{anisur.rahman@saha.ac.in}
\affiliation{Durgapur Govt. College, Durgapur - 713214, Burdwan, West Bengal, INDIA}

\date{\today}
\begin{abstract} A model where a Dirac fermion is coupled to
 background dilaton
field is considered to study s-wave scattering of fermion by a
back ground dilaton black hole. It is found that an uncomfortable
situation towards information loss scenario arises when one loop
correction gets involved during bosonization.
\end{abstract}
\maketitle
\section{introduction}
Scattering of fermion off dilaton black hole has been  extensively studied
over the years  \cite{HAW, HOOFT, GARF, STRO, SUS, GIDD, GIDD1,
 CALL,  MIT, ARINF} and it has provided much insight into its connection to the
  Hawking radiation. Even after the intensive investigations it remains as a
  subject of several interests because of the subtleties involved in it in
  connection with information loss scenario during the formation
  and subsequent evaporation of black hole. It is worth mentioning at this
  stage that a controversy in this context was generated from
  the Hawking's suggestion \cite{HAW} three decades ago.
  However his recent suggestion on this issue \cite{HAW1} has brought
   back a pretty pleasant scenario. It may even be thought that the
   controversy has come to an end.

 General description of such scattering
 problem is extremely difficult. Despite that, there have been attempts
 in studying such problem in its
full complexity through the  s-matrix description of such event
\cite{HOOFT}. Comparatively less complicated model therefore
entered into  the picture and showed its prominent role in this
issue \cite{STRO, SUS, GIDD, GIDD1, CALL}.  The toy model due to
Alford and Strominger \cite{STRO} provides an interesting
description of the s-wave scattering of fermion off dilaton black
hole with a trustworthy results concerning information loss. It
helps to avoid some of the technical obstacle posed by quantum
gravity in $(3+1)$. Even in the presence of gravitational anomaly
a systematic description of this scattering of fermion off dilaton
would have been possible through this model \cite{MIT, ARINF}. The
provision of taking the effect of one loop correction \cite{JR,
RABIN} into consideration is also an exiting aspect of this model.
That indeed shows a way to investigate the effect of anomaly
\cite{MIT, ARINF} on this scattering phenomena. Notably, this
model arose in two dimensional non-critical string theory and its
black hole solution was discovered in \cite{WAD}.

  Few years back Mitra studied this scattering problem replacing Dirac fermion by chiral fermion and found
  an uncomfortable scenario \cite{MIT}. He observed that information failed to be preserved.
  With the use
  of anomaly we have shown that that disaster can be avoided \cite{ARINF}. Anomaly played there a very
  surprising as well as interesting role. Seeing the interesting role of anomaly on the s-wave scattering
  of chiral fermion \cite{MIT, ARINF} we are intended here to investigate the role of one loop correction on
   the s-wave
  scattering of Dirac fermion using the said toy model due to Alford and Strominger
  \cite{STRO}. Needless to mention that the counter term appeared
  here due to one loop correction looks similar to the term used
  in \cite{ARINF}.

\section{Two dimensional effective model for studying s-wave scattering}
Here we consider only a special case which turns out to be particularly
simple scattering of s-wave fermion incident on dilaton black hole.
Black hole is the extrema of the following (3+1) dimensional action.
\begin{equation}
S_{AF} = \int d^4 x\sqrt{-g}[R + 4(\nabla\phi)^2
- {1\over 2}F^2 + i \bar\psi D\!\!\!/\psi]. \label{EQ1}
\end{equation}
Here $g$ represents determinant of the space time metric.
The geometry consists of three regions \cite{STRO}.
Far from the black hole there is an asymptotically flat region.
The mouth leads to an infinitely long throat region.
In side the throat region the metric is approximated by the
flat metric on two dimensional Minkowsky space
times the round metric on two sphere with radius $Q$.
Electromagnetic field strength is tangential over
the two sphere and an integer to
$4\pi Q$. When the energy scales is large compared to $Q$, the radius
  of the two sphere the dynamics within the throat region can be
  described by the effective action
\begin{equation}
S_{AF} = \int d^2\sigma\sqrt{-g}[R + 4(\nabla\phi)^2 + {1\over
{Q^2}} - {1\over 2}F^2 + i \bar\psi D\!\!\!/\psi], \label{EQ2}
\end{equation}
Here $D_\mu=\partial_\mu +
eA_\mu$.
It is a two dimensional effective field theory of dilaton
gravity coupled to fermion.  $\Phi$ represents the scalar
dilaton field and $\psi$ is the charged fermion.
 For sufficiently low
energy incoming fermion, gravitational effect on the scattering of s-wave fermion incident
on a charge dilaton black hole can be neglected
and equation (\ref{EQ2}) can be approximated by
\begin{equation}
{\cal S}_f = \int d^2x[i\bar\psi\gamma^\mu[\partial_\mu +
ieA_\mu]\psi - {1\over 4} e^{-2\Phi(x)}F_{\mu\nu}F^{\mu\nu}]].
\label{EQ3}
\end{equation}

The coupling $e$ has one mass dimension. The indices $\mu$ and
$\nu$ takes the values $0$ and $1$ in $(1+1)$ dimensional space
time. The dilaton field $\Phi$ stands as a non dynamical back
ground. It completes its role here just by making the coupling
constant a position dependent function. This very toy model of
quantum gravity in $(1+1)$ dimension  contains black holes and
Hawking radiation in a significant manner. Let us now define
$G^2(x)= e^{2\Phi(x)}$. We will choose a particular dilaton
background motivated by the linear dilaton vacuum of $(1+1)$
dimensional gravity like the other standard cases \cite{GIDD,
GIDD1, CALL, STRO, SUS, MIT, ARINF}. Therefore, $\Phi(x) = -x^1$,
where $x^1$ is space like coordinate. The region $x^1 \to\ +
\infty$, corresponds to exterior space where the coupling $G^2(x)$
vanishes and the fermion will be able to propagate freely.
However, the region where $x^1 \to -\infty$, the coupling constant
will diverge and it is analogous to infinite throat in the
interior of certain magnetically charged black hole.

Equation
(\ref{EQ2}), was derived viewing the throat region of a four
dimensional dilaton black hole as a compactification from four to
two dimension \cite{GARF, GIDD, STRO}. Note that, in the extremal
limit, the geometry is completely non-singular and there is no
horizon but when a low energy particle is thrown into the
non-singular extremal black hole, it produces a singularity and an
event horizon. The
geometry of the four dimensional dilaton black hole consists of
three significant regions \cite{GARF, STRO, SUS, GIDD, GIDD1} as has
already been  mentioned.  As long as
one proceed nearer to the black hole the curvature begins to rise
and finally enters into the mouth region (the entry region to the
throat).  In the deep throat region physics will be governed by
the equation (\ref{EQ2}) since the metric at that region gets simplified into
flat two dimensional Minkowsky metric times the round metric
on the two sphere with radius $Q$.
The dilaton field $\Phi$ indeed increases linearly with the proper
distance into the throat.

We are now in a state to start our analysis and we would like to
proceed with the bosonized version of the theory (\ref{EQ3}).
During the process of Bosonization a one loop correction
automatically enters within the action because bosonization needs
to integrate out both the the left handed as well as the right
handed part of the fermion one by one that leads to a fermionic
determinant \cite{JR, AR, AR1}. When this fermionic determinant is
expressed in terms of scalar field a one loop correction enters
into the theory in order to remove the divergence of the fermionic
determinant. So the tree level bosonized theory gets the effect of
loop correction during the process. Of course, bosonization can be
done keeping the gauge symmetry intact which was used in
\cite{STRO}. Here masslike term for gauge field has been taken
into consideration since we are intended to study the effect of
this one loop correction in the s-wave scattering of Dirac
fermion. With the counter term used in the study of non confining
Scgwinger model \cite{AR, AR1} the bosonized action reads
\begin{equation}
{\cal L}_B = {1\over 2}\partial_\mu\phi \partial^\mu\phi -
e\tilde\partial_\mu\phi A^\mu + {1\over 2}ae^2A_{\mu}A^{\mu} -
{1\over 4}e ^{2\Phi(x)}F_{\mu\nu}F^{\mu\nu}. \label{LBH}
\end{equation}
Here $\phi$ represents a scalar field and $\tilde\partial_\mu$ is
the dual to $\partial_\mu$. $\tilde\partial_\mu$ is defined by
$\tilde\partial_\mu=\epsilon_{\mu\nu}\partial^\nu$. Note that the lagrangian
(\ref{LBH}), maps onto the
non-confining Schwinger model \cite{AR, AR1} for $\Phi(x)=0$,.

 The $U(1)$ current in this
situation is
\begin{equation}
J_\mu = -e\epsilon_{\mu\nu}\partial^\nu\phi + ae^2A_\mu
\end{equation}
and it is non conserving since $\partial_\mu J^\mu \neq 0$. This
current was of preserving nature in \cite{GIDD, GIDD1, STRO, SUS}
and in those situations the currents were $J_\mu=
-e\epsilon_{\mu\nu}\partial^\nu\phi$. The new setting considered
here indeed to show the role of the one loop correction on the
s-wave scattering of Dirac fermion.
\section{Hamiltonian analysis of the model}
It is now necessary to carry out the Hamiltonian analysis of the
theory to observe the effect of the dilaton field on
the equations of motion. From the standard definition the
canonical momenta corresponding to the scalar field $\phi$, and
the gauge fields $A_0$ and $A_1$ are found out:
\begin{equation}
\pi_\phi = \phi' - eA_1\label{MO1}
\end{equation}
\begin{equation}
\pi_0 = 0,\label{MO2}
\end{equation}
\begin{equation}
\pi_1 = e^{-2\phi(x)}(\dot A_1 - A_0')={1\over {G^2}}(\dot A_1 -
A_0').\label{MO3}
\end{equation}
Here $\pi_\phi$, $\pi_0$ and $\pi_1$ are the momenta corresponding
to the field $\phi$, $A_0$ and $A_1$. Using the above equations,
it is straightforward to obtain the canonical hamiltonian through
a Legendre transformation. The canonical hamiltonian is found out
to be
\begin{eqnarray}
{\cal H} &=& {1\over 2}(\pi_\phi +eA_1)^2 + {1\over
2}e^{2\Phi(x)}\pi_1^2 + {1\over 2}\phi'^2 + \pi_1A_0' -eA_0\phi' \nonumber\\ &-&
{1\over 2}ae^2(A_0^2 - A_1^2).\label{CHAM}\end{eqnarray}
Note that
there is an explicit space dependence in the hamiltonian (\ref{CHAM})
through the dilaton field $\Phi(x)$ but it does not pose any hindrance
to be preserved in time. So consistency and physically sensibility are in no way be threatened.
Equation (\ref{MO2}) is the familiar
primary constraints of the theory. Therefore, it is necessary to
write down an  effective hamiltonian:
\begin{equation}
{\cal H}_{eff} = {\cal H}_C + u\pi_0
\end{equation}
where $u$ is an arbitrary Lagrange multipliers. The primary
constraints (\ref{MO2}) has to be preserve in order to have a
consistent theory. The preservation of the constraint (\ref{MO2}),
leads to the Gauss law of the theory as a secondary constraint:
\begin{equation}
G = \pi_1' + e\phi' +  ae^2A_0 \approx 0. \label{GAUS}
\end{equation}
The preservation of the constraint (\ref{GAUS}) though does not
give rise to any new constraint it fixes the velocity $u$ which
comes out to be
\begin{equation}
u =A'_1. \label{VEL}
\end{equation}
We, therefore, find that the phase space of the theory contains
the following two second class constraints.
\begin{equation}
\omega_1 = \pi_0 \approx 0, \label{CON1}
\end{equation}
\begin{equation}
\omega_2 = \pi_1' + e\phi' +  ae^2A_0 \approx 0.\label{CON2}
\end{equation}
Both the constraints (\ref{CON1}) and (\ref{CON2}) are weak
conditions up to this stage. When we impose these constraints
strongly into the canonical hamiltonian (\ref{CHAM}), the
canonical hamiltonian gets simplified into the following form.
\begin{eqnarray}
{\cal H}_{red} &=& {1\over 2}(\pi_\phi + eA_1)^2 + {1\over
{2ae^2}}(\pi'_1 + e\phi')^2 + {1\over 2}e^{2\Phi(x)}\pi_1^2
\nonumber\\
&+&{1\over 2}\phi'^2 + {1\over 2}ae^2A_1^2.
\label{RHAM}\end{eqnarray}
$H_{red}$ obtained in equation
(\ref{RHAM}), is generally known as reduced Hamiltonian. According
to Dirac \cite{DIR}, Poisson bracket gets invalidate for this reduced
Hamiltonian. This reduced Hamiltonian however remains
consistent with the Dirac bracket which is defined by
\begin{eqnarray}
& &[A(x), B(y )]^* = [A(x), B(y)] \nonumber \\
&-&\int[A(x), \omega_i(\eta)]
C^{-1}_{ij}(\eta, z)[\omega_j(z), B(y)]d\eta dz, \label{DEFD}
\end{eqnarray}
where $C^{-1}_{ij}(x,y)$ is given by
\begin{equation}
\int C^{-1}_{ij}(x,z) [\omega_j(z), \omega_k(y)]dz =\delta(x-y)
\delta_{ik}. \label{INV}
\end{equation}
For the theory under consideration \noindent $C_{ij}(x,y)=$
\begin{equation}
ae^2 \pmatrix {0 & -\delta(x-y) \cr \delta(x-y) & o \cr}.
\label{MAT}
\end{equation}

Here $i$ and $j$ runs from $1$ to $2$ and $\omega$'s represent the
constraints of the theory. With the definition (\ref{DEFD}), we
can compute the Dirac brackets between the fields describing the
reduced Hamiltonian $H_{red}$. The Dirac brackets between the
fields $A_1$, $\pi_1$, $\phi$ and $\pi_\phi$ are required to
obtain the theoretical spectra (equations of motion):
\begin{equation}
[A_1(x), A_1(y)]^* = 0 = [\pi_1(x), \pi_1(y)]^* ,\label{DR1}
\end{equation}
\begin{equation}
[A_1(x), \pi_1(y)]^* = \delta(x-y),\label{DR2}
\end{equation}
\begin{equation}
[\phi(x), \phi(y)]^* = 0 =[\pi_\phi(x), \pi_\phi(y)]^* ,\label{DR3}
\end{equation}
\begin{equation}
[\phi(x), \pi_\phi(y)]^* = \delta(x-y). \label{DR4}
\end{equation}
The  Dirac Brackets (\ref{DR1}), (\ref{DR2}), (\ref{DR3}) and
(\ref{DR4}), along with the Heisenberg's equation of motion leads
to the following four first order equations.
\begin{equation}
\dot A_1= e^{2\Phi}\pi_1 -{1\over {ae^2}}(\pi_1'' + e\phi'')
,\end{equation}

\begin{equation}
\dot\phi = \pi_\phi + eA_1 ,
\end{equation}

\begin{equation}
\pi_\phi = {{a+1}\over a}\phi'' + {1\over {ae}}\pi_1''
,\end{equation}

\begin{equation}
\dot\pi_1 = -e\pi_\phi - (a+1)e^2A_1. \end{equation}
A little
algebra converts the four first order equations into the following
two second order Klein-Gordon equations:

\begin{equation}
[\Box + (1+a)e^2e^{2\Phi(x)}]\pi_1 = 0, \label{SP1}
\end{equation}
\begin{equation}
\Box[\pi_1 + e(1+a)\phi] = 0. \label{SP2}
\end{equation}

The equation (\ref{SP1}), represents a massive boson with square
of the mass $m^2 =(1+a)e^2e^{2\Phi(x)}$. Here $a$ must be greater
than $-1$ in order to have the mass of the boson a physical one.
Equation (\ref{SP2}), however describes a massless boson. The
presence of this massless boson has a disastrous role here which
will be going to uncovered now.

Let us concentrate into the theoretical spectra. Equation (\ref{SP1}) shows that the
 mass of the boson is not constant in this model.
It contains a position dependent factor $G^2=e^{2\Phi(x)}$, where $\Phi = -x^1$,
for the background generated by the linear
dilaton vacuum of $(1+1)$ dimensional gravity. Therefore, $m^2 \to
\ + \infty$ when $x^1 \to\ - \infty$ and $m^2 \to \ 0$ when $x^1
\to\ + \infty$. Thus mass of the boson goes on increasing
indefinitely in the negative $x^1$ direction which implies that
any finite energy contribution must be totally reflected and an
observer at $x^1 \to \infty$ will get back all the  information. To be
more precise, mass will vanish near the mouth (the entry region to
the throat) but increases indefinitely as one goes into the throat
because of the variation of this space dependent factor $G^2$.
Since massless scalar is equivalent to massless fermion in $(1+1)$
dimension, we can conclude that a massless fermion proceeding into
the black hole will not be able to travel an arbitrarily long
distance and will be reflected back with a unit probability and a unitary
 s-matrix can be constructed. So
there is no threat regarding information loss from the massive
sector of the theory. Of course it is a pleasant scenario. However
an uncomfortable situation appears when we observe carefully
towards the massless sector of the theory (\ref{SP2}).
It will remain massless irrespective of its position because
unlike the massive sector it does not contain any space dependent
factor. So this fermion will be able to travel within the black
hole without any hindrance and an observer at $x^1 \to \infty$
will never find this fermion with a backward journey. Thus a real
problem towards information loss appears for this setting. Note
that in the similar type of studies, \cite{STRO, SUS, GIDD, GIDD1}
where the setting was such that the masslike term for gauge field
was absent, this problem did not occur. The result of the present
work though leads to an uncomfortable situation, there is no known
way to avoid it. It is true that after the Hawking's recent
suggestion \cite{HAW1} it seems to be an unwanted and
untrustworthy scenario but one can not rule it out too if he has
to accept the model \cite{STRO}. More serious investigation is
needed indeed
 in this issue. It is true that this result indicates a
 less brighter side of the model but it's presence can not be ignored or suppressed.

\section{Conclusion}
In this letter the s-wave scattering of fermion off dilaton black
hole is investigated in presence of one loop correction due to
bosonization. It was found that the presence of that correction
term brings a disastrous result. Information loss could not be
avoided. The result was not in agreement with the Hawking's recent
suggestion too. But there is no way to rule out this possibility.

Role of this type of quantum correction due to bosonization does
not come as a great surprise for the first time. The crucial role
of that was noticed earlier in the description of quantum
electrodynamics and quantum chiral electrodynamics \cite{AR, AR1,
BEL, JR, KH, PM, MG, FLO}. A famous instance in this context is
the removal of the long suffering of the chiral electrodynamics
from the non unitary problem \cite{JR}.

\noindent{\bf Acknowledgment}: It is a pleasure to thank the
Director, Saha Institute of Nuclear Physics and the Head of the
Theory Group of Saha Institute of Nuclear Physics, Kolkata for
providing working facilities. I would like to thanks the referee
for his suggestion toward the improvement of this manuscript.

\end{document}